\documentclass{elsart}

\usepackage{graphics}
\usepackage{graphicx}

\usepackage{amssymb}

\begin{document}

\begin{frontmatter}



\title{Time-Evolution of Collective Meson Fields and Amplification of 
Quantum Meson Modes in Chiral Phase Transition}


\author{Yasuhiko Tsue}
\ead{tsue@cc.kochi-u.ac.jp}

\address{Physics Division, Faculty of Science, Kochi University, 
Kochi 780-8520, Japan}

\begin{abstract}
The time evolution of quantum meson fields in the O(4) linear sigma model is 
investigated in a context of the dynamical chiral phase transition. 
It is shown that amplitudes of quantum pion modes are amplified 
due to both mechanisms of a parametric resonance and a resonance by the 
forced oscillation according to the small oscillation of the chiral condensate 
in the late time of chiral phase transition. 
\end{abstract}
\vspace{-3mm}
\end{frontmatter}

\def\mib#1{\mbox{\boldmath $#1$}}
\newcommand{\bra}[1]{\langle {#1} |}     
\newcommand{\ket}[1]{| {#1} \rangle}     
\vspace{-5mm}
\section{Introduction}

\vspace{-5mm}

One of the recent interests associated with the physics of the 
relativistic heavy-ion collisions is to study the dynamics of matter at 
very high energy density. Especially, it is interesting to 
investigate the dynamics of the chiral phase transition in connection with 
the problem of the formation of a disoriented chiral condensate (DCC). 
In Fig.1, the schematic diagram of the chiral phase transition is depicted 
in the $O(4)$ linear sigma model. 
\begin{figure}[t]
  	\begin{center}
	\includegraphics[width=1.0\linewidth]{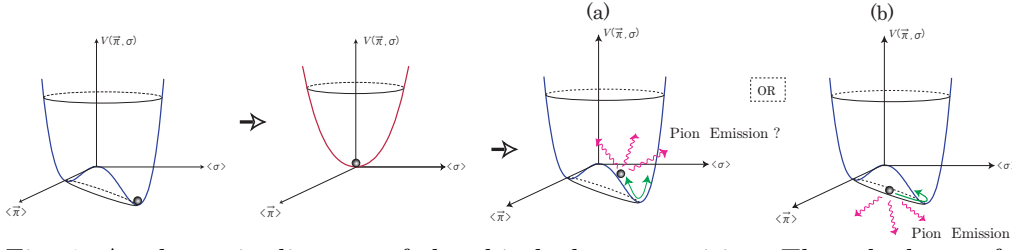}
	\end{center}
	\vspace{-4mm}
\caption{\small A schematic diagram of the chiral phase transition. 
The role-down of the condensate is realized in (a).
The collective isospin rotation is realized in (b). 
}
\end{figure}
In order to investigate the time-evolution of the order 
parameter and of the fluctuation modes around it 
in the chiral phase transition, we have formulated the time-dependent 
variational approach to dynamics of quantum fields in terms of 
a squeezed state or a Gaussian wave functional\cite{TVM}. 
The main advantage of this approach to the dynamical problem 
lies in the fact that both the degrees of freedom of the mean field and the 
quantum fluctuations can be treated self-consistently. 
By using the above-mentioned method, 
we investigated the dynamics of a collective isospin rotation 
of quantum meson fields corresponding to the situation 
which is schematically drawn in Fig.1(b) 
in the $O(4)$-linear sigma model\cite{TVM}. 
We concluded that the collisionless dissipation occurs and we obtained that 
the damping time is about 40 fm/$c$ for the energy density 
(160 MeV)$^4$. Further, the number of emitted mesons is estimated, 
which is about 15 pions per fm/$c$ in this dissipative process\cite{TVM}. 

In this paper, it is assumed that the chiral phase transition 
occurs by the rolling down of the chiral condensate in what is called 
quench scenario, which is schematically depicted in Fig.1(a). 
The time evolution of chiral phase transition is investigated in 
this situation. 
In the early time of this process, the spinodal 
decomposition may occur. On the other hand, it is interesting in this paper 
to the mechanism of the chiral phase transition in the late time. 
Many authors pointed out that the parametric amplification of 
pion modes with low momenta is realized\cite{MM,HM}, which leads to the pion 
emission. Here, we will point out that the resonance mechanism 
due to forced oscillation is also realized as well as the parametric 
resonance.

\vspace{-3mm}

\section{Parametric Resonance versus Forced Oscillation}
\vspace{-3mm}
Let us start with the following Hamiltonian density 
of the $O(4)$ linear sigma model :
\vspace{-0.3cm}
\begin{equation}\label{1}
    \mathcal{H} =
      \frac{1}{2} \pi_{a}({\mib x})^2 
      +\frac{1}{2} \nabla \phi_{a}({\mib x}) \cdot\nabla\phi_{a}({\mib x})
      +\lambda\left(\phi_{a}({\mib x})^2-{m^2}/{4\lambda} \right)^2 
         -H\phi_{0}({\mib x}) .
\end{equation}
\vbox{}
\vspace{-0.8cm}
The squeezed state is adopted as a trial state in the time-dependent 
variational principle : 
\vspace{-0.4cm}
\begin{eqnarray}\label{2}
\ket{\Phi(t)}
=N\exp\left(i({\overline \pi}\!\cdot\!\phi
-{\overline\varphi}\!\cdot\!\pi)\right)
\exp(\phi\cdot\![-(G^{-1}-G^{(0)}{}^{-1})/4
+i\Sigma]\!\cdot\!\phi)\ket{0} , \quad
\end{eqnarray}
where $N$ represents a normalization factor and 
${\overline \pi}\cdot\phi=\sum_{a=0}^3\int\! d^3\!{\mib x}
{\overline \pi}_a({\mib x},t)\phi_a({\mib x})$ and 
$\phi\cdot G^{-1}\cdot\phi=\sum_{a=0}^3\int\!\int\! d^3\!{\mib x}d^3\!{\mib y}
\phi_a({\mib x})G_{a}^{-1}({\mib x},{\mib y},t)\phi_a({\mib y})$ 
and so on. Here, $G^{(0)}_a=\bra{0}\phi_a({\mib x})\phi_a({\mib y})\ket{0}$. 
The expectation values of the field operators are obtained as, 
for example, 
$\bra{\Phi(t)}\phi_a({\mib x})\ket{\Phi(t)}
={\overline \varphi}_a({\mib x},t)
$
and 
$\bra{\Phi(t)}\phi_a({\mib x})\phi_a({\mib y})\ket{\Phi(t)}=$\break
${\overline \varphi}_a({\mib x},t){\overline \varphi}_a({\mib y},t)
+G_a({\mib x},{\mib y},t)
$. 
Thus, ${\overline \varphi}_a({\mib x},y)$ and $G_a({\mib x},{\mib y},t)$ 
represent the mean field and the fluctuation around it, respectively. 
The two-point function $G_a({\mib x},{\mib y},t)$ can be further expanded 
in terms of the mode functions under the assumption of the translational 
invariance : 
$G_{a}({\mib x},{\mib y},t)
=\int_{\bf k}
e^{i{\bf k}\cdot({\bf x}-{\bf y})}
\eta_{\bf k}^{a}(t)^2
$, 
where $\int_{\bf k}=\int d^3{\bf k}/{(2\pi)^3}$. 
The time-dependence of these functions, together with ${\overline \pi}_a$ and 
$\Sigma_a$, are determined in the time-dependent variational principle 
which can be expressed as 
$\delta\int_{t_1}^{t_2}dt \bra{\Phi(t)}i{\partial}/{\partial t}
-{\hat H}\ket{\Phi(t)}=0$, 
where ${\hat H}=\int d^3{\mib x}\ {\mathcal H}$.

The equations of motion for the condensate ${\overline \varphi}_a(t)$ 
and the fluctuation modes 
$\eta_{\bf k}^a(t)$ under the translational invariance can be 
expressed as 
\vspace{-0.6cm}
\begin{eqnarray}\label{3}
  & &{\ddot {\overline \varphi}}_{a}(t)
     -m^2 {\overline \varphi}_{a}(t) 
     + 4\lambda {\overline \varphi}_{a}(t)^3
     +12\lambda \int_{\bf k}\eta_{\bf k}^{a}(t)^2\cdot 
       {\overline \varphi}_{a}(t) \nonumber\\
  & & \qquad\qquad\qquad
     +4\lambda \sum_{b\neq a}\left({\overline \varphi}_{b}(t)^2 
     +\int_{\bf k}\eta_{\bf k}^{b}(t)^2 \right)
     {\overline \varphi}_{a}(t) -H\delta_{a0} =0 \ , \nonumber\\
  & & {\ddot \eta}_{\bf k}^{a}(t) 
    + \biggl[{\mib k}^2-m^2+12\lambda {\overline \varphi}_{a}(t)^2 
    + 12\lambda\int_{{\bf k}'}\eta_{{\bf k}'}^{a}(t)^2 
       \nonumber\\
  & & \qquad\qquad\qquad
     +4\lambda \sum_{b\neq a}\left({\overline \varphi}_{b}(t)^2 
     +\int_{{\bf k}'}\eta_{{\bf k}'}^{b}(t)^2 
     \right) 
       \biggl]
    \eta_{\bf k}^{a}(t) 
    -{1}/{4 \eta_{\bf k}^{a}(t)^3} = 0 \ . \qquad
\end{eqnarray}
\begin{figure}[b]
  	\begin{center}
	\includegraphics[width=0.32\linewidth]{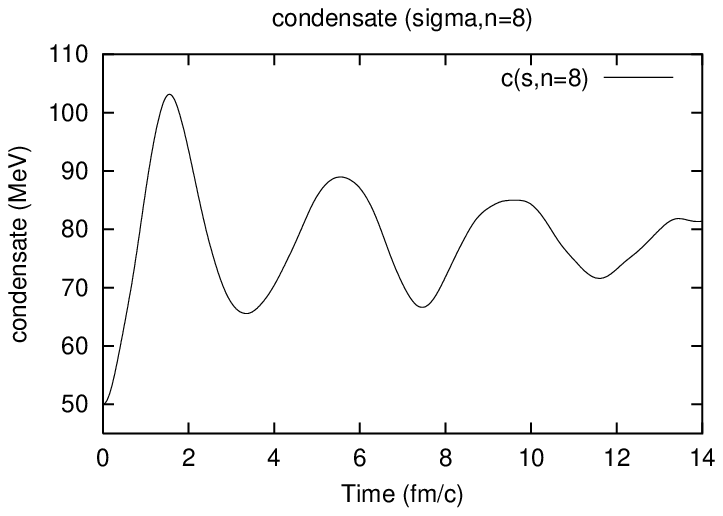}
	\includegraphics[width=0.32\linewidth]{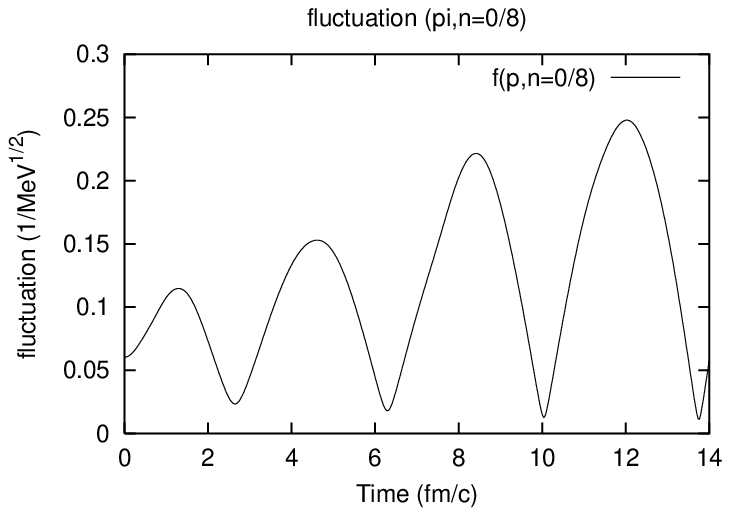}
	\includegraphics[width=0.32\linewidth]{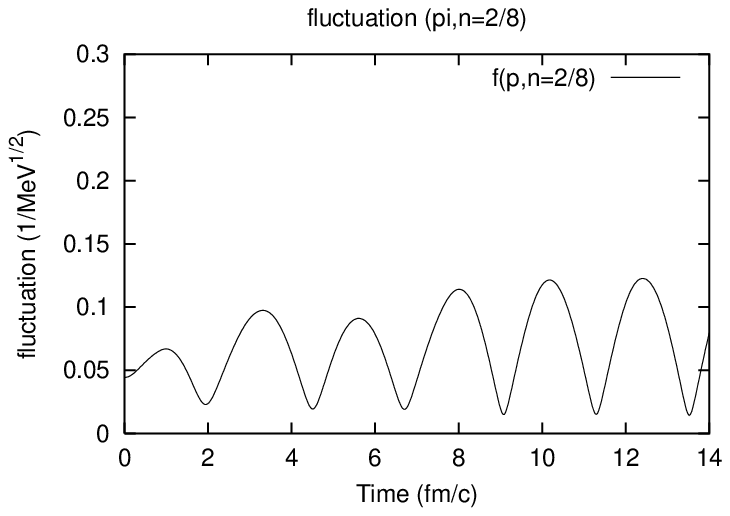}
	\end{center}
\vspace{-2mm}
\caption{\small The time-evolutions of the mean field (left) and the lowest 
(middle) and the second excited pion modes (right).}
\end{figure}
The time-evolution of the mean field\cite{TKI}, i.e., the chiral condensate 
with the direction of the sigma field, is depicted in the left in Fig.2. 
The relaxation to the vacuum value is seen due to the inclusion of 
the quantum fluctuation modes. If we do not take into account the quantum 
fluctuation modes fully, the relaxation is not seen. 
Thus, we conclude that the 
quantum fluctuation modes play an important role in this relaxation 
process. 
Also, the amplification of the amplitudes of quantum fluctuation modes 
with low momenta in the direction of pion fields occurs accompanying with the 
relaxation of the chiral order parameter\cite{TKI}, which is seen in 
the middle and the right figures in Fig.2. 
This phenomena may be understood in 
terms of a parametric amplification as was mentioned by many authors. 
However, there is another possibility to amplify the fluctuation 
modes\cite{T}. 

In the late time of the chiral phase transition, the dynamical variables 
can be expanded around the static configurations. Namely, 
${\overline \varphi}_0(t)=\varphi_0+\delta\varphi(t)$ and 
$\eta_{\bf k}^a(t)=\eta_{\bf k}^a+\delta\eta_{\bf k}^a(t)$, 
where $\varphi_0$ and $\eta_{\bf k}^a$ are static solutions of (\ref{3}). 
In this expansion, we assume that 
${| \delta\varphi(t) |}/{\varphi_0} \ll 1$ and 
$|\delta\eta_{\bf k}^a(t)/\eta_{\bf k}^a| \ll 1$. 
Then we obtain the approximate equations of motion for the condensate 
and fluctuation modes as 
\vspace{-4mm}
\begin{eqnarray}
& &\delta{\ddot \varphi}(t)+M_\sigma^2 \delta\varphi(t) = 0 \ , 
\quad \hbox{\rm i.e.,}\quad
\delta\varphi(t) =\delta\sigma \cos(M_\sigma t) \ , 
\label{6}\\
& &\delta{\ddot \eta}_{\bf k}^{\sigma}(t)
+\left[4({\mib k}^2+M_\sigma^2)+24\lambda\varphi_0 \delta\varphi(t)\right]
\delta\eta_{\bf k}^\sigma(t) 
= -24\lambda\eta_{\bf k}^{\sigma} \varphi_0 \delta\varphi(t) \ , 
\nonumber\\
& &\delta{\ddot \eta}_{\bf k}^{\pi}(t)
+\left[4({\mib k}^2+M_\pi^2)+8\lambda\varphi_0 \delta\varphi(t)\right]
\delta\eta_{\bf k}^\pi(t) 
= -8\lambda\eta_{\bf k}^{\pi} \varphi_0 \delta\varphi(t) \ , 
\label{7}
\end{eqnarray}
where $\eta_{\bf k}^0$ ($\eta_{\bf k}^i,\ i=1,2,3$) and 
$\delta\eta_{\bf k}^0(t)$ ($\delta\eta_{\bf k}^i(t),\ i=1,2,3$) have been 
rewritten as $\eta_{\bf k}^\sigma$ ($\eta_{\bf k}^\pi$) 
and $\delta\eta_{\bf k}^\sigma(t)$ ($\delta\eta_{\bf k}^\pi(t)$), 
respectively. 
Here, $M_\sigma$ ($M_\pi$) represents the sigma (pi) meson mass 
in the $O(4)$ linear sigma model, respectively. 
If the right-hand sides of (\ref{7}) can be neglected or 
have no effects, these equations give amplifying solutions in a certain 
parameter regions 
due to the parametric resonance mechanism. Actually, the amplitude 
of the lowest pion mode is amplified as is seen in the middle of Fig.2 
in the late time of chiral phase transition. 
Even if the parametric resonance is not realized, there is another 
possibility to make amplitudes of meson modes amplify. 
Actually, the amplitude of the second excited pion mode is 
amplified due to the forced oscillation mechanism. In our numerical 
values, 
\vspace{-3mm}
\begin{equation}\label{30}
\delta{\eta}_{n=2}^{\pi}(t)
\approx 0.71 (M_\sigma/2\sqrt{2M_\pi})\cdot(\delta\sigma/\varphi_0) \cdot
t \sin (2\sqrt{{\mib k}_2^2+M_\pi^2}\ t) \ ,
\end{equation}
\vspace{-0.6cm}
where ${\mib k}_2^2=(2\pi/L)^2\times 2$ for the second excited 
mode under the box normalization of the length $L(=\!10$ fm). 
This behavior is seen in the right of~Fig.2.

\vspace{-3mm}
\section{Conclusion}
\vspace{-3mm}
It has been shown approximately and analytically that both the parametric 
amplification and the resonance due to the forced oscillation occur 
in the late time of the chiral phase transition.
Thus, it is pointed out that there actually coexist two mechanisms to amplify 
the quantum fluctuation modes in the late time of 
the chiral phase transition, namely, 
that a forced oscillation works as well as a parametric resonance.

\vspace{-3mm}



\end{document}